\documentclass[aps,twocolumn,superscriptaddress]{revtex4-2}  

\usepackage{graphicx}  
\usepackage{dcolumn}   
\usepackage{bm}        
\usepackage{amssymb}   
\usepackage{amsmath}
\usepackage{scalerel}
\usepackage{braket}
\usepackage{siunitx}

\begin{document}

\title{Observation of Full Hierarchy of Temporal Quantum Correlations with a Superconducting Qubit}
\date{\today}

\author         {Hao-Cheng Weng}
 \email          {haocheng.weng@bristol.ac.uk}
 \affiliation    {Department of Physics and Center for Quantum Science and Technology, National Tsing Hua University, Hsinchu 30013, Taiwan}
 \affiliation    {Quantum Engineering Technology Labs and H. H. Wills Physics Laboratory, University of Bristol, Bristol BS8 1TL, United Kingdom}
 
\author         {Chen-Yeh Wei}
 \affiliation    {Department of Physics and Center for Quantum Science and Technology, National Tsing Hua University, Hsinchu 30013, Taiwan}  

\author         {Huan-Yu Ku} 
\affiliation{Department of Physics, National Taiwan Normal University, Taipei 11677, Taiwan}
 
\author         {Shin-Liang Chen}
\affiliation{Department of Physics, National Chung Hsing University, Taichung 402, Taiwan}
\affiliation{Physics Division, National Center for Theoretical Sciences, Taipei 10617, Taiwan}
 
\author         {Yueh-Nan Chen}
 \affiliation    {Department of Physics and Center for Quantum Frontiers of Research \& Technology (QFort), National Cheng Kung University, Tainan 701, Taiwan}
 
\author         {Chih-Sung Chuu}
 \email          {cschuu@phys.nthu.edu.tw}
 \affiliation    {Department of Physics and Center for Quantum Science and Technology, National Tsing Hua University, Hsinchu 30013, Taiwan} 

\begin{abstract}

Temporal quantum correlations provide an intriguing way of testing quantumness at the macroscopic level, with a logical hierarchy present among the quantum correlations associated with nonmacrorealism, temporal steering, and temporal inseparability. By manipulating the dynamics of a superconducting qubit, we observe the full hierarchy of temporal quantum correlations. Moreover, we show that the rich dynamics of the temporal quantum correlations, such as sudden death or revival of temporal steering, provides a useful and unique measure for benchmarking qubits on a realistic circuit. Our work finds applications in identifying the casual structure in a quantum network, the non-Markovianity of open quantum systems, and the security bounds of quantum key distribution. As an example, we demonstrate the non-Markovianity of a single superconducting qubit on the quantum circuit.

\end{abstract}

\maketitle

\section{Introduction}
Temporal correlations of a quantum system provide a brilliant way of testing the quantum nature at the macroscopic level, with the most celebrated example being the Leggett-Garg inequality \cite{Leggett85,Emary13,GiuseppePRA2023}. The Leggett-Garg inequality, often regarded as Bell's inequality~\cite{Bell64} in time, examines whether the two-time correlation of a system satisfies the macroscopic realism and noninvasive measurability. If the latter can be ruled out (for example, by weak continuous measurements~\cite{Ruskov06}), the violation of the inequality then reveals the quantumness of the system \cite{Palacios-Laloy10,Waldherr11,Goggin11,Dressel11,Knee12,Robens15,Ku20,Lucas2024arXiv,HsiehPRL2025,Hsieh2025PRA}. However, in much the same way that not all entangled states can violate the Bell inequality \cite{Werner89}, not every quantum system can demonstrate the two-time correlation required to violate the Leggett-Garg inequality. Such an incompleteness, which is common to all measures of the temporal quantum correlations, arises from the types of quantum state characterization used to quantify the temporal correlations. As a result, there is a logical hierarchy between the temporal quantum correlations associated with nonmacrorealism, temporal steering, and temporal inseparability \cite{Fritz10,YNChen14,Fitzsimons15,Ku18}. 

The hierarchy of the temporal quantum correlations not only attracts fundamental interests but also finds applications in quantum information tasks. For example, it can be exploited to identify casual structures in a quantum network \cite{Ried15,Ku18}, the non-Markovianity of open quantum systems \cite{SLChen16}, metrologic test \cite{LeePRR2023} or the security bounds of the quantum key distribution with trusted or untrusted devices \cite{YNChen14,Bartkiewicz16}. However, the observation of the full hierarchy of temporal or spatial quantum correlations still remains challenging, with the major obstacle being the decoherence suffered by quantum systems. In this Article, by manipulating the dynamics of a superconducting qubit, we experimentally demonstrate the full hierarchy of the temporal quantum correlations. Each level of the hierarchy is unambiguously ranked by the sudden death of nonmacrorealism, temporal steering, and temporal inseparability on distinct time scales. Moreover, we show that the rich dynamics of the temporal quantum correlations, such as the sudden death or the revival of temporal steering (a signature of non-Markovianity), provides a useful and unique measure for benchmarking the qubits on a realistic circuit.

\begin{figure}[t]
\includegraphics[scale=1]{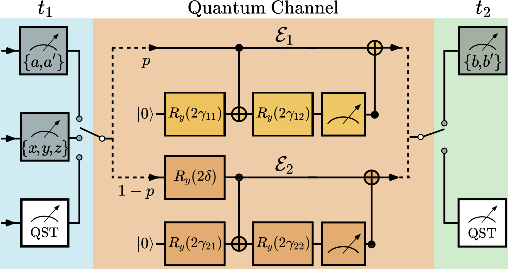}
\caption{\label{fig:1} Quantum circuit for testing and manipulating the temporal quantum correlations. The measurements at time $t_1$ and $t_2$ are selected accordingly (see text) when the nonmacrorealism, temporal inseparability, and temporal steering are tested. The quantum channel is composed of two extreme channels with unequal probability. With the probability and rotation gates properly chosen (see text), the amplitude-damping, dephasing, and depolarizing channels can be implemented. When the qubits are left freely, the quantum channel is replaced by recurrent identity gates.}
\end{figure}

\section{Temporal quantum correlations}
To quantify different temporal quantum correlations, we prepare a superconducting qubit in a maximally mixed state initially and study the temporal correlation of the qubit states before (at time $t_1$) and after (at time $t_2$) evolving on the quantum circuit (Fig.~\ref{fig:1}). The use of a maximally mixed state ensures that the measurement at time $t_1$ does not change the statistics at a later time $t_2$, that is, $\sum_{a=\pm1}p(a,b) = p(b) \ \forall \  \hat{\sigma}_a$--the so-called no-signaling in time condition. The types of the state characterization at time $t_1$ and $t_2$ rigorously define different measures of the temporal quantum correlations. For quantifying the nonmacrorealism, the measurements at time $t_1$ and $t_2$ are $\hat{\sigma}_{m_1}\in\{\hat{\sigma}_a, \hat{\sigma}_{a'}\}$ and $\hat{\sigma}_{m_2} \in \{\hat{\sigma}_b, \hat{\sigma}_{b'}\}$, respectively. With the measurement outcomes $\{a, a'\} \in \pm 1$ and $\{b, b'\} \in \pm 1$, the measure is given by 
\begin{equation}
 B_{\rm max} = {\max}\bigg\{ 0,\frac{C_{ab}+C_{a'b}+C_{ab'}-C_{a'b'}-2}{2\sqrt{2}-2} \bigg\},
 \label{eq:1}
\end{equation}
where $C_{ab}$ is the expectation value for obtaining the outcome $ab$. In our experiment, $\hat{\sigma}_a, \hat{\sigma}_{a'}, \hat{\sigma}_b$, and $\hat{\sigma}_{b'}$ are chosen to be $\left(\hat{\sigma}_y+\hat{\sigma}_z\right)/\sqrt{2}, \left(-\hat{\sigma}_y+\hat{\sigma}_z\right)/\sqrt{2}, \hat{\sigma}_z$, and $\hat{\sigma}_y$, respectively, so that $B_{\rm max}$ reaches the maximum 1 when $C_{ab}+C_{a'b}+C_{ab'}-C_{a'b'}$ maximally violates the temporal CHSH inequality \cite{Fritz10, ku2021hidden} and gives the minimum 0 when the measurement outcomes obey the macrorealistic theory. In Appendix \ref{AppendixB}, we compare our method with the quasiprobability approach \cite{halliwell2016leggett} and show that the same no-signaling in time condition holds for both cases.

As for the temporal steering measure, we employ the temporal steering robustness (TSR),
\begin{equation}
 {\rm TSR}=\min\left(\mathrm{Tr}\sum\limits_{\lambda} \sigma_\lambda -1 \right),\textrm{ with } \sigma_\lambda \geq 0 \quad \forall \lambda,
\end{equation}
subject to
\begin{equation}
\sum\limits_{\lambda} p(a|m_1,\lambda) \sigma_\lambda - p(a|m_1) \sigma_{a|m_1} (t_2) \geq 0\quad  \forall a,m_1,
\end{equation}
which was introduced in \cite{Ku18} but has not been explored experimentally. The TSR searches for any hidden-state model \cite{KuPRA2016,Ku18,SLChen16,SLChen17} with pre-determined quantum states $\sigma_\lambda$ and probability distribution $p(a|m_1,\lambda)$ that resemble the states $\sigma_{a|m_1} (t_2)$ with probability $p(a|m_1)$ obtained by the quantum state tomography (QST) at time $t_2$, given the outcome $a$ at time $t_1$. If no hidden-state model can be found (TSR $\neq 0$), the system is temporally steerable provided that the no signaling in time is obeyed or, equivalently, the initial state is prepared in a maximally mixed state \cite{Ku18}. Experimentally, we take the measurement at time $t_1$ to be $\hat{\sigma}_{m_1} \in \{\hat{\sigma}_x, \hat{\sigma}_y, \hat{\sigma}_z\}$.

Lastly, to quantify the temporal inseparability, the quantum state tomography are performed at both $t_1$ and $t_2$ to obtain the psuedodensity matrix,
\begin{equation}
R=\frac{1}{4}\sum\limits_{i,j} C_{ij} \cdot \sigma_i \otimes \sigma_j,
\end{equation}
where $C_{ij}$ is the expectation values of these measurements in the $\sigma_i \otimes \sigma_j$ basis and $\sigma_i \in \{\hat{\mathbb{I}},\hat{\sigma}_x,\hat{\sigma}_y,\hat{\sigma}_z\}$. The psuedodensity matrix is not positive semidefinite if it cannot be written as a product of the states at time $t_1$ and $t_2$ (temporally inseparable). Thus, the $f$-function \cite{Fitzsimons15,Leggett80},
\begin{equation}
f = \parallel R \parallel _{\rm tr} -1,
\end{equation}
which gives 0 when the psuedodensity matrix is positive semidefinte, can be used to quantify how temporally inseparable the quantum system is.

\begin{figure}[t]
\includegraphics[scale=0.9]{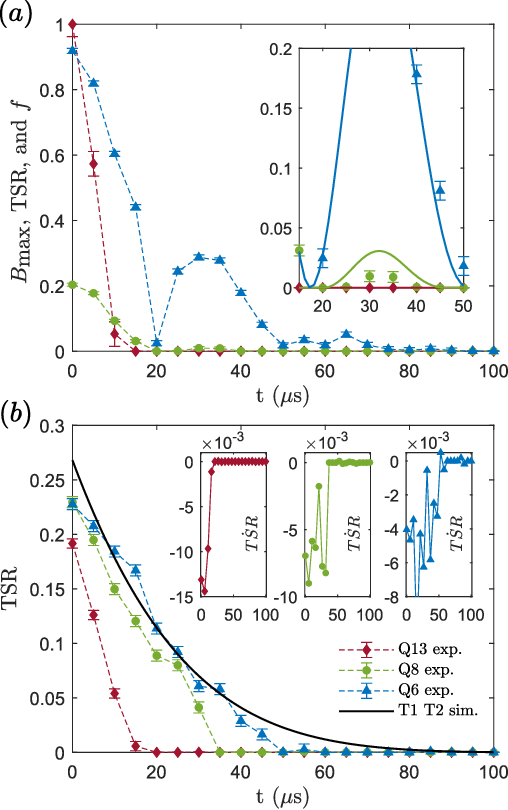}
\caption{\label{fig:2} (a) The measured $B_{\rm max}$ (diamond dots), TSR (circular dots), and \textit{f}-function (triangle dots) of a qubit left freely on the quantum circuit. The nonmacrorealism ($B_{\rm max} > 0$), temporal steering (TSR $\neq 0$), and temporal inseparability ($f \neq 0$) are observed. Non-Markovianity is also evident with the revival of the temporal steering robustness (see also the inset). The curves are the calculated $B_{\rm max}$, TSR, and \textit{f}-function, where $\gamma_A/\hbar J=0.224$ and $\gamma_P/\hbar J=0.038$ for the system qubit and $\gamma_A/\hbar J=0.359$ and $\gamma_P/\hbar J=0.083$ for the environment qubit. (b) The measured TSR of three different qubits left freely on the quantum circuit. The curve is the calculated TSR assuming an isolated qubit subject to the amplitude damping and dephasing. The insets show the corresponding instantaneous rates.}
\end{figure}

Using the measures introduced above, we first characterize the temporal correlations of the qubits left freely on the quantum circuit. To do this, the identity gates (gate time $\sim 0.1\ \mu$s) are repeatedly applied to the qubits in between the correlation measurements of $B_{\rm max}$ (diamond dots), TSR (circular dots), and \textit{f}-function (triangle dots). An example of a qubit from the IBM quantum computing system (qubit No.8, ibmq\_16\_melbourne) is shown in Fig.~\ref{fig:2}(a). The nonmacrorealism ($B_{\rm max} > 0$), temporal steering (TSR $\neq 0$), and temporal inseparability ($f \neq 0$) are clearly evident. A variety of dynamics are also observed among different temporal quantum correlations. For example, while the temporal inseparability (triangle dots) exhibits an oscillating behavior before damping out, the nonmacrorealism (diamond dots) suddenly vanishes around $t=15\ \mu$s--a phenomena conceptually related to the sudden death of entanglement \cite{Almeida07,Yu09}. 

\begin{figure}[t]
\includegraphics[scale=1]{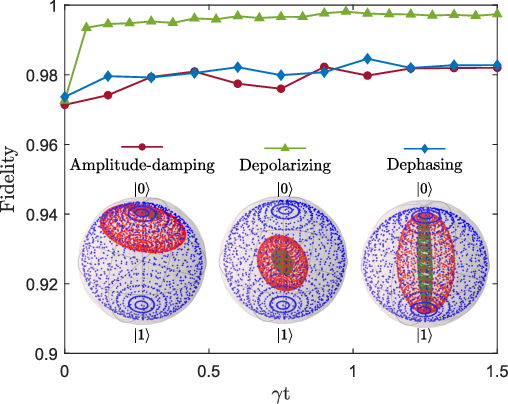}
\caption{\label{fig:3} Fidelity of the quantum channels, obtained by quantum process tomography, versus time. The inset illustrates the geometric interpretation of a maximally mixed state (blue dots) evolving over time in the amplitude-damping, depolarizing, and dephasing channels (from left to right). The red and green dots represent the quantum states at two later times in order.}
\end{figure}

The temporal steerability (circular dots) also suffers the sudden death (TSR = 0) but it then revives before fading away. We attribute the revival to the non-Markovianity \cite{SLChen16}, in which the environment surrounding the quantum system has the ``memory'' of its past evolution and allows the ``information'' to flow back and forth between the quantum system and environment. In our experiment, the non-Markovianity likely results from the crosstalk between the qubit in study and the other idle qubits. To verify this, we consider a ``system'' qubit interacting with an ``environment'' qubit in the presence of the amplitude damping and dephasing. Taking the interaction $H_{\rm int}=\hbar J(\sigma_1^+ \sigma_2^-+\sigma_1^- \sigma_2^+)$, the amplitude-damping rate $\gamma_A \simeq 1/ \rm T_1$, and the dephasing rate $\gamma_P \simeq 1/\rm T_2-1/2 \rm T_1$, where $T_1$ and $T_2$ are the longitudinal and transverse relaxation times of the system qubit, the simplified model (curves in the inset) reproduces the revival and oscillation. The sudden death of the temporal steerability can also provide useful information for benchmarking the qubits. For example, Fig.~\ref{fig:2}(b) shows the measured TSR of three different qubits (No.6, 8, 13 of ibmq\_16\_melbourne) left freely on the same quantum circuit. All three qubits suffer the sudden death of temporal steerability (see also the instantaneous rates in the insets) but in very different time scales. Both the occurrence and timing of the sudden death cannot be simply described by considering an isolated qubit subject to the amplitude damping and dephasing (curve).

\section{Quantum channels}
Owing to the decoherence on the quantum circuit, it is crucial to control the quantum dynamics of the qubits in order to resolve the hierarchy of the temporal quantum correlations. We thus implement several quantum channels on the circuit, including the amplitude-damping channel, dephasing channel, and depolarizing channel. All of these quantum channels can be decomposed into two extreme channels \cite{Lu17}, 
\begin{eqnarray}
&&\mathcal{E}_1=
\left( \begin{array}{cc} \cos{\beta} & 0 \\
0 & \cos{\alpha} \end{array} \right), \\
&&\mathcal{E}_2=
\left( \begin{array}{cc}0 & \sin{\alpha} \\
\sin{\beta} & 0 \end{array} \right) R_y(2\delta),
\end{eqnarray}
with probability $p$ and $1-p$, respectively. These extreme channels can be implemented by the CNOT gates and sequential rotation gates $R_y(2\gamma_{j1})$, $R_y(2\gamma_{j2})$, and $R_y(2\delta)$ (Fig.~\ref{fig:1}) with $2\gamma_{j1,j2}=\beta \mp \alpha \pm \pi/2$. By controlling $p$, $\gamma_{j1}$, $\gamma_{j2}$, and $\delta$, we experimentally realize the amplitude-damping channel ($p=1, \gamma_{11}=-\gamma_{12}, \delta=0$), depolarizing channel ($\gamma_{11}=\pi/4, \gamma_{21}=\pi/4, \gamma_{22}=0, \delta=\pi/2$), and dephasing channel ($\gamma_{11}=-\gamma_{12}=\pi/4, \gamma_{21}=-\gamma_{22}=-\pi/4, \delta=0$). The inset of Fig.~\ref{fig:3} illustrates the geometric interpretation of these quantum channels on a Bloch sphere. To verify the fidelity of these quantum channels $\mathcal{E}(\rho)=\sum_{mn} \tilde{\mathcal{E}}_m \rho \tilde{\mathcal{E}}_n^{\dagger} \chi_{mn}$, where $\tilde{\mathcal{E}}_{m,n} \in \{ \hat{\mathbb{I}}, \hat{\sigma}_x,-i\hat{\sigma}_y,\hat{\sigma}_z \}$, we also carry out the quantum process tomography \cite{Nielsen11}. The fidelity of the tomographically reconstructed $\chi_{mn}$ as shown in Fig.~\ref{fig:3} is high for all quantum channels, with slight deviations resulting from the nonideal gate operations.

\section{Hierarchy of temporal quantum correlations}
\begin{figure}[t]
\centering
\includegraphics[scale=0.72]{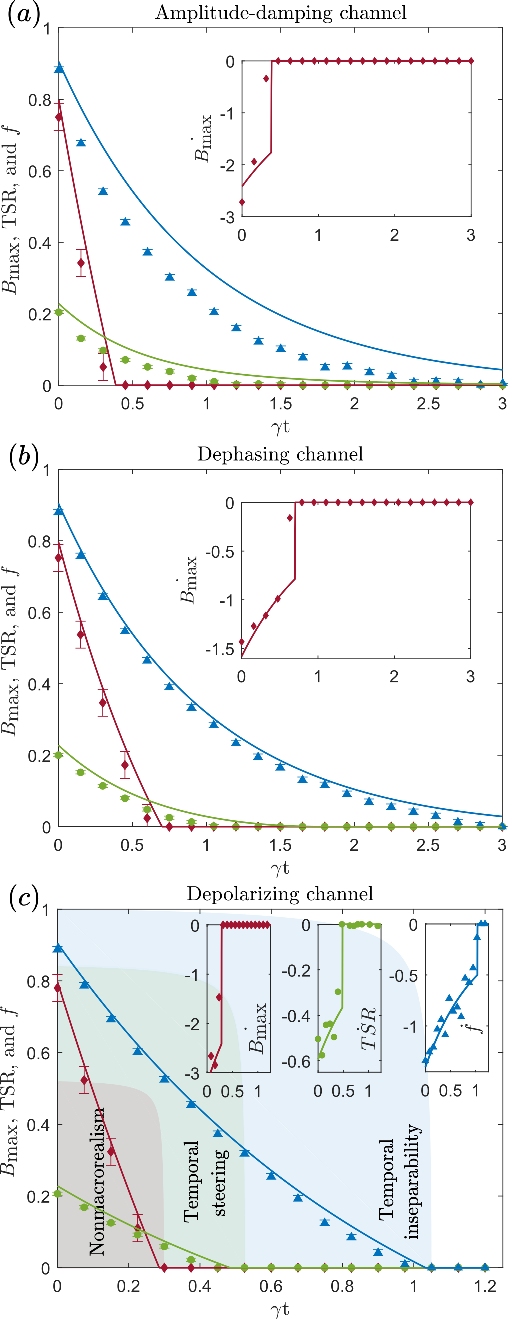}
\caption{\label{fig:4} The measured (dots) and calculated (curves) $B_{\rm max}$ (diamonds), TSR (circles), and \textit{f}-function (triangles) versus $\gamma t$ in the (a) amplitude-damping channel, (b) dephasing channel, and (c) depolarizing channel. $\gamma$ are the corresponding decoherence rates. The full hierarchy of the temporal quantum correlations is observed in the depolarizing channel, where all quantum correlations experience the sudden death. The calculation uses the master equations in Lindblad form. The insets show the instantaneous rates of the temporal quantum correlations exhibiting the sudden death.}
\end{figure}

With the controllable quantum channels implemented on the circuit, we next study the quantum dynamics of a single qubit in these channels. The target qubit (qubit No.0, ibmq\_ourense) and control qubit (qubit No.1, ibmq\_ourense) are chosen so that the $T_1$ and $T_2$ times are long compared to the total time needed for the state initialization, quantum channel, and measurements. Note that the full circuit can be completed within 2 $\mu$s (estimated from the operation time), which is much shorter than the time scale required for any non-Markovian noise to take effect (around 20 $\mu$s as observed in Fig.~\ref{fig:2}). Fig.~\ref{fig:4} shows the measured $B_{\rm max}$ (diamond dots), TSR (circular dots), and \textit{f}-function (triangle dots) versus $\gamma t$ in the (a) amplitude-damping, (b) dephasing, and (c) depolarizing channels, with the corresponding decoherence rates denoted by $\gamma$. The observed dynamics of the nonmacrorealism ($B_{\rm max} > 0$), temporal inseparability (TSR $\neq 0$), and temporal steerability ($f \neq 0$) in different channels are in good agreement with the calculation (curves) by the master equations in Lindblad form \cite{Nielsen11,Ku18} and the known decoherence ($T_1$ and $T_2$ times) of the qubits. The discrepancy between our observation and calculation possibly comes from the nonideal two-qubit gates.

Of interest to this work is the temporally resolvable sudden deaths of different temporal quantum correlations. As shown in Fig.~\ref{fig:4}, the temporal quantum correlations and their instantaneous rates (insets) cease to zero abruptly in one or more quantum channels. In contrast to the amplitude-damping and dephasing channels [Fig.~\ref{fig:4}(a) and \ref{fig:4}(a), respectively], where only the sudden death of the nonmacrorealism ($B_{\rm max} = 0$) is observed, the depolarizing channel [Fig.~\ref{fig:4}(c)] is intriguing that all temporal quantum correlations suffer the sudden death in the same channel. Moreover, the distinct timings of these sudden deaths allow us to reconstruct the logical hierarchy of different temporal quantum correlations. For example, as $\gamma t > 0.3$, the qubit still exhibits the temporal steerability and inseparability when it obeys the macrorealistic theory. Overall, the hierarchy indicates that a nonmacrorealistic system is temporally steerable and inseparable (but not vice versa) while a temporally steerable system is temporally inseparable (but not vice versa).

\section{Conclusion}
We have demonstrated the full hierarchy of the nonmacrorealism, temporal inseparability, and temporal steerability in a superconducting quantum circuit. This is accomplished by manipulating the decoherence of the qubits to induce temporally resolvable sudden death of different temporal quantum correlations in a single quantum channel. The hierarchy can be exploited to explore the casual structures of a quantum network \cite{Ried15,Ku18,KuPRXQ2022,ChangPRR2024} or the security bounds of the quantum key distribution with trusted and untrusted devices \cite{YNChen14,Bartkiewicz16}. The distinct dynamics of the temporal steerability observed with different qubits also provides useful information for benchmarking the qubits. By characterizing the qubits with the temporal steering robustness, we observed the revival of the temporal steerability--a signature of the non-Markovianity that cannot be characterized by conventional qubit performance metrics. The non-Markovian process can benefit the quantum communication \cite{Shounak18,Huelga12,Laine14} in which the sudden death of the temporal quantum correlations may lead to the loss of quantum information. Our work demonstrates that hierarchy is a powerful and experimentally easy-to-implement tool to explore the underlying physics of noisy intermediate-scale quantum machines.

\begin{acknowledgments}
We acknowledge the use of IBM Quantum services for this work. Access to IBM Q systems is provided by the IBM Q Hub at National Taiwan University, Taiwan. HCW, CYW, and CSC acknowledge the support of the National Science and Technology Council, Taiwan (Grant No. 113-2119-M-007-012). YNC acknowledges the support of the Ministry of Science and Technology, Taiwan (MOST Grants No. 107-2628-M-006-002-MY3 
and No. MOST 109-2627-M-006-004), and the U.S. Army Research Office (ARO Grant No. W911NF-19-1-0081). HYK is supported by the Ministry of Science and Technology, Taiwan, (Grants No.~MOST 112-2112-M-003 -020 -MY3), and Higher Education Sprout Project of National Taiwan Normal University (NTNU) and the Ministry of Education (MOE) in Taiwan.
\end{acknowledgments}

\appendix

\section{Calculating the \textit{f}-function}
To derive Eq. 5, we start from the formulation in Eq. 4. Equation 4 is the general form for pseudodensity matrix and conventional bipartite density matrix, both having unit trace (sum of eigenvalues are 1). However, for conventional density matrix, the $R$ is positive semi-definite (all eigenvalues are positive) while $R$ permits negative eigenvalues in case of pesudodensity matrix, which then implies temporal correlation as discussed in \cite{Fitzsimons15}. In Eq. 5, $\|R\|_{tr}$ is the sum of all singular values. Since $R$ is a Hermitian matrix, the absolute values are exactly the singular values. As a result, subtracting $\|R\|_{tr}$ by one gives the contribution of the negative eigenvalues. On the other hand, we denote that psuedodensity matrix is not positive semidefinite if it cannot be written as a product of the states at time $t_1$ and $t_2$ (temporally inseparable). This is according to \cite{Ku18} in which the following proposition is made
\begin{equation}
f=0 \  \Rightarrow \ \exists \ \xi_\lambda, \zeta_\lambda, \ \ \textrm{such that} \ R=\sum_\lambda p(\lambda) \ \xi_\lambda \otimes \zeta_\lambda.
\end{equation}
Here, $p(\lambda)$ is a probability distribution and $\xi_\lambda$ and $\zeta_\lambda$ are some valid quantum states at $t_1$ and $t_2$.

Here, we give an example on calculating the \textit{f}-function. Consider the scenario in Fig.~\ref{fig:1} where the quantum channel is an identity channel. We have the pseudodensity matrix
\begin{equation}
    R=\left( 
    \begin{array}{cccc}
    1 & 0 & 0 & 0 \\
    0 & 0 & 0.5 & 0 \\
    0 & 0.5 & 0 & 0\\
    0 & 0 & 0 & 0 \\
     \end{array} \right),
\end{equation}
whose eigenvalues are \{{$\pm$1/2,0,1\}. $\|R\|_{tr}$ sums all singular values (absolute values of the eigenvalues), giving us $f=\|R\|_{tr}-1=2-1=1$.

\section{Comparison with the quasiprobability approach}
\label{AppendixB}
In this work, nonmacrorealism is discussed through violation of the temporal CHSH inequality, where the expectation value $C_{ab}$ as in Eq.~(\ref{eq:1}) is given by
\begin{equation}
\begin{split}
    C_{ab} & =p(a=1, b=1)+p(a=-1,b=-1) \\
    & -p(a=1,b=-1)-p(a=-1,b=1)
\end{split}
\end{equation}
and the probability $p(a,b)$ is
\begin{equation}\label{eqs2.2}
p(a,b)  =tr(\hat{\sigma}_b \sqrt{\hat{\sigma}_a} \rho_0 \sqrt{\hat{\sigma}_a})
\end{equation}
with $\rho_0$ being the initial state. Since the maximally mixed state is considered throughout our work, $\rho_0=\frac{\mathcal{I}}{2}$ and $\mathcal{I}\sqrt{\hat{\sigma}_a}=\sqrt{\hat{\sigma}_a}\mathcal{I}$, the probability becomes
\begin{equation}
\begin{split}
    p(a,b) & =tr(\hat{\sigma}_b \sqrt{\hat{\sigma}_a} \frac{\mathcal{I}}{2}\sqrt{\hat{\sigma}_a})  \ \   
    \\
    & =tr(\hat{\sigma}_b \sqrt{\hat{\sigma}_a}\sqrt{\hat{\sigma}_a} \frac{\mathcal{I}}{2}) \ \   
    \\
    & = \frac{1}{2}tr(\hat{\sigma}_b \hat{\sigma}_a),
\end{split}
\end{equation}
Compared with the quasiprobability approach \cite{halliwell2016leggett}, in which nonmacrorealism is characterized by the quasiprobability $q(a,b)$,
\begin{equation}\label{eqs2.2}
    q(a,b)  =\frac{1}{2}tr\{[\hat{\sigma}_b \hat{\sigma}_a +\hat{\sigma}_a\hat{\sigma}_b]\rho_0\}, 
\end{equation}
the probability and quasiprobability are equivalent. This is because, when $\rho_0=\frac{\mathcal{I}}{2}$, the quasiprobability becomes
\begin{equation}
\begin{split}
    q(a,b) & =\frac{1}{2}tr\{[\hat{\sigma}_b \hat{\sigma}_a +\hat{\sigma}_a\hat{\sigma}_b]\frac{\mathcal{I}}{2}\}  \   
    \\
    & =\frac{1}{2}[tr(\hat{\sigma}_b \hat{\sigma}_a \frac{\mathcal{I}}{2})+ tr(\hat{\sigma}_a\hat{\sigma}_b\frac{\mathcal{I}}{2})]\\
    & = \frac{1}{2}[tr(\hat{\sigma}_b \hat{\sigma}_a \frac{\mathcal{I}}{2})+ tr(\hat{\sigma}_a \frac{\mathcal{I}}{2} \hat{\sigma}_b)] \ \   
    \\
    & = \frac{1}{2}[tr(\hat{\sigma}_b \hat{\sigma}_a \frac{\mathcal{I}}{2})+ tr(\hat{\sigma}_b \hat{\sigma}_a \frac{\mathcal{I}}{2})] \\  
    & = \frac{1}{2}tr(\hat{\sigma}_b \hat{\sigma}_a),
\end{split}
\end{equation}
where we use $\hat{\sigma}_b\mathcal{I}=\mathcal{I}\hat{\sigma}_b$ and the cyclic property of trace.
As the quasiprobability approaches readily satisfy one of the weakest no-signaling in time conditions called linear positivity by Goldstein and Page \cite{goldstein1995linearly}, the same no-signaling in time condition also holds with nonmacrorealism characterized by the standard probabilities used in our work.

\bibliographystyle{unsrt} 
\bibliography{Paper}

\end{document}